\documentstyle[12pt,epsf,epsfig,rotate]{article}

\newcommand \be	   {\begin{equation}}
\newcommand \ee    {\end{equation}}
\newcommand \lan   {\langle}
\newcommand \ran   {\rangle}
\newcommand \si	   {\sigma_{i}}
\newcommand \sj	   {\sigma_{j}}
\newcommand \Tc    {\mbox{$T_{c}$}}
\newcommand \dd    {{\rm d}}
\newcommand \ra    {\rightarrow}
\newcommand \lra   {\longrightarrow}

\begin{document}

\title{Equilibrium and off-equilibrium simulations of the 4$d$
Gaussian spin glass}

\author{Giorgio Parisi, Federico Ricci Tersenghi and
Juan J. Ruiz-Lorenzo\\[0.5em]
{\small Dipartimento di Fisica and Infn, Universit\`a di Roma}
{\small {\em La Sapienza}}\\
{\small P.le A. Moro 2, 00185 Roma (Italy)}\\[0.3em]
{\small \tt parisi@roma1.infn.it}\\
{\small \tt ricci@chimera.roma1.infn.it}\\
{\small \tt ruiz@chimera.roma1.infn.it}\\[0.5em]}

\date{May, 1996}

\maketitle

\begin{abstract}

In this paper we study the on and off-equilibrium properties of the
four dimensional Gaussian spin glass.

In the static case we determine the critical temperature and the
critical exponents with more precision that in previous simulations.

In the off-equilibrium case we settle the general form of the
autocorrelation function, and show, for the first time, that is
possible to obtain, dynamically, a value for the order parameter.

\end{abstract}  

\thispagestyle{empty}
\newpage

\section{\protect\label{S_INT}Introduction}

Nowadays, the problem of the full characterization of the phase
transition in finite dimensional spin glasses is still opened both
from the static and dynamical approaches.

Our discussion is focused on the four dimensional case (the same
applies in the more physical case, the three dimensional system).

The equilibrium (static) simulations show a very neat intersection of
the Binder cumulant curves, that is a signal of a phase transition at
finite temperature with an order parameter (the Edward-Anderson order
parameter, that we will denote hereafter as $q_{EA}$).
We can identify this order parameter with the position of a Dirac
delta in the probability distribution of the overlap, $P(q)$.
Up to now, both the SRSB theory (Spontaneous Replica Symmetry
Breaking)\cite{ME_PA_VI,GIORGIO} and the droplet theory \cite{DROPLET}
are compatible with this result.
Differences concern the shape of the rest of the $P(q)$. In the
droplet theory $P(q)$ is the sum of two Dirac deltas, one in $q_{EA}$
and another in the opposite overlap, and has a Binder cumulant equal
to 1. The SRSB theory maintains this structure too, but adds a
continuous non zero part in the interval $(-q_{EA},q_{EA})$.
This is a non trivial distribution that has a Binder cumulant
different from 1, except at $T=0$ where the SRSB theory predicts two
pure states like the droplet theory.

The main problem is the impossibility of a direct measure of the order
parameter, $q_{EA}$.
The scaling of the peak of $P(q)$ seems compatible both with a
Kosterlitz-Thouless (KT) transition, i.e. $q_{peak} \sim
1/L^{\alpha}$, and with a scaling like $q_{peak} = q_{EA}
+a/L^{\rho}$. Obviously the KT scenario goes against the intersection
of the Binder cumulant curves.
A possible explanation of this phenomena could be
that the term $a/L^{\rho}$ is bigger than $q_{EA}$ for the range of
lattice sizes that has been simulated and hence the latter is
unobservable. Simulation of  bigger lattices should be done in order
to get  $q_{EA} \gg a/L^{\rho} $.

The dynamical approach \cite{CU_KU} has the same problematic of the
previously discussed static case.
The main physical quantity in this approach is the spin-spin
autocorrelation, defined as:
\be
C(t,t_w)=\frac1N\sum_{i=1}^N\overline{\lan\si(t_w)\si(t_w+t)\ran}
\;\;,
\ee

Usually in the literature \cite{RIEGER_REV} one  finds the empirical
formula
\begin{equation}
C(t,t_w)=t^{-x} f(t/t_w) \;\;,
\label{E_corre}
\end{equation}
for instance in the Mean Field case \cite{CU_KU_RI} and in the three
dimensional case \cite{RIEGER_93}\footnote{ Also in an on-equilibrium
numerical simulation in the three dimensional case \cite{OGIEL}: $C(t)
\sim t^{-x}$.}.

The static limit (on-equilibrium situation) is achieved sending $t_w$
to infinite first, and then simulating  larger  $t$. The formula
(\ref{E_corre}) for the spin-spin autocorrelation function goes to
zero in this limit. However, in the case of a non zero order
parameter, this autocorrelation function must  go to $q_{EA}$.
It is clear that in the regime of $t_w \gg t \gg 1$ should be found a
formula like 
\begin{equation}
C(t,t_w)=(q_{EA}/f(0)+a t^{-x}) f(t/t_w) \;\;,
\label{corre}
\end{equation}
but a very long numerical simulation is needed in order to observe
both the terms $q_{EA}$ and $a t^{-x}$.
In the present work we show for the first time numerical evidence of
this kind of behaviour.

Up to now, the only numerical studies of off-equilibrium dynamics in
finite dimensional spin-glass are those of H. Rieger \cite{RIEGER_93}
in the three dimensional case.
 
The four dimensional case seems easier to simulate, since it is far
away from the lower critical dimension of the spin glasses ($d_l<3$
\cite{MA_PA_RI_RU_95,KAWA_YOUNG}) and the static is thus very clear.
In this paper we will study mainly the off-equilibrium dynamics of
this model in order to compare with the three dimensional results by
Rieger and to examine the  possibility to extract a finite value for
the order parameter. In addition simulations  have been performed in
the static (on-equilibrium) case in order to characterize with higher
precision the location of the critical temperature and the critical
exponents.

Both for the off and on-equilibrium cases we review the numerical
results from the point of view of the previous discussion and we  try
to link both approaches in order to obtain a conclusion regarding the
existence of a finite temperature phase transition, with a non zero
order parameter.

\section{\protect\label{S_static}Model, simulation and static
observables}

We have studied the 4-$d$ Ising spin glass with nearest neighbor
interactions and zero external magnetic field, whose Hamiltonian is
\be
{\cal H}= -\sum_{<i,j>} J_{ij}\sigma_i\sigma_j \;\; ,
\ee
where $<\!i,j\!>$ denotes nearest neighbor pairs and the couplings are
extracted from a Gaussian distribution with zero mean and unit
variance.

The static and dynamical behaviour of the model have been investigated
by several different simulations during which many
different observables have been measured .
This section describes the way we performed measurements of the static
exponents and the critical temperature with a precision higher than
that available in the literature \cite{BHATT_YOUNG, BADONI}.

The equilibrium simulations have been performed on small lattices ($L
= 3, 4, 5, 6, 7, 8$) to ensure the system reached equilibrium.
Most of the work has been made in a range of temperature around the
critical one, \Tc.
The average over the disorder, has been done on 2048 samples for all
the lattice sizes.
For each realization of the quenched disorder we have simulated two
replicas with spin $\sigma_i$ and $\tau_i$.
This enabled us to measure the $k$-th cumulant of the distribution of
the overlaps, $q^{(k)} \equiv \int q^k P(q) \dd q$, simply by
averaging the quantity $\left ( N^{-1} \sum_i\sigma_i\tau_i\right)^k$
over a large number of independent configurations.

All the calculation have been carried on a {\em tower} of the parallel
supercomputer APE100 \cite{APE100}, with a real performance of about 5
Gigaflops.

A detailed study has been devoted to the calculation of the number of
sweeps needed to reach the equilibrium and to the estimate of the
autocorrelation time at the equilibrium.
This study suggests a thermalization time of about $10^5$ sweeps, being
sure that using this value even the biggest system at the lowest
temperature will be thermalized.
To verify the correctness of this value we studied the evolution of
the biggest system ($L=8$) at the lowest temperature ($T=1.7$): we
choose three replicas of the system such that having, at the starting
time, two overlaps set to zero and the third one equal to one; we have
followed the evolution of these overlaps averaging over a large number
of disorder configurations and we have estimated the thermalization
time as the time needed in order that the three overlaps converge to a
single value.

Once the equilibrium has been reached, we  measured how much time
was needed to decorrelate  the observables. Particularly, we have
seen that the overlap between two replicas has a time correlation
function that decreases exponentially, $C(t) \sim \exp (-t/\tau)$.
This defines a characteristic time whose typical value at $T=1.7$
are: for $L=4 \;\;\; \tau \sim 200$, for $L=6 \;\;\; \tau \sim 1000$
and for $L=8 \;\;\; \tau \sim 3000$.
In the final simulation, after thermalization, we measured every $\tau$
sweeps the overlap between the two replicas for a time longer than the
equilibration one.

\begin{figure}
\begin{center}
\leavevmode
\centering\epsfig{file=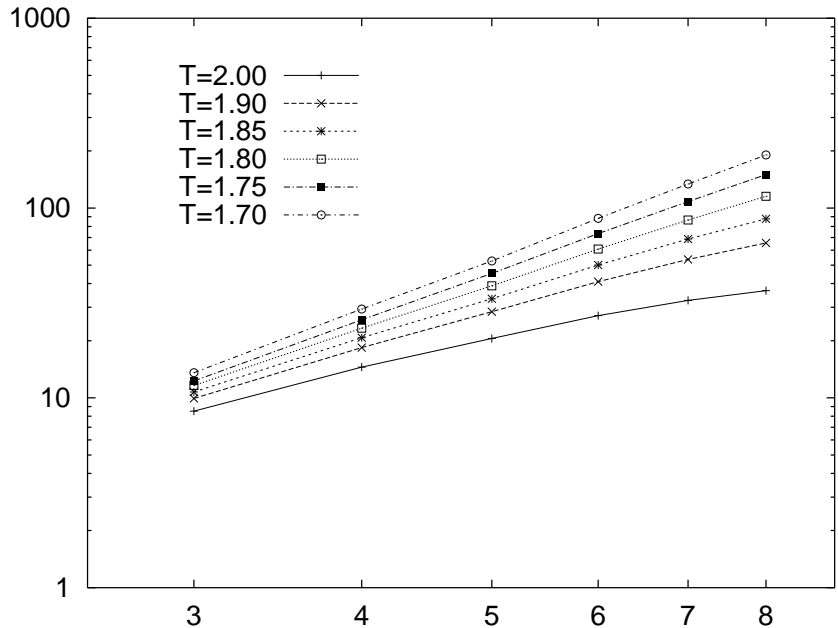,width=0.75\linewidth}
\end{center}
\protect\caption{$\chi_{SG}(L,T)$ vs. $L$; the errors are of the
order of the symbol. \protect\label{F_chi}}
\end{figure}

\begin{figure}
\begin{center}
\leavevmode
\centering\epsfig{file=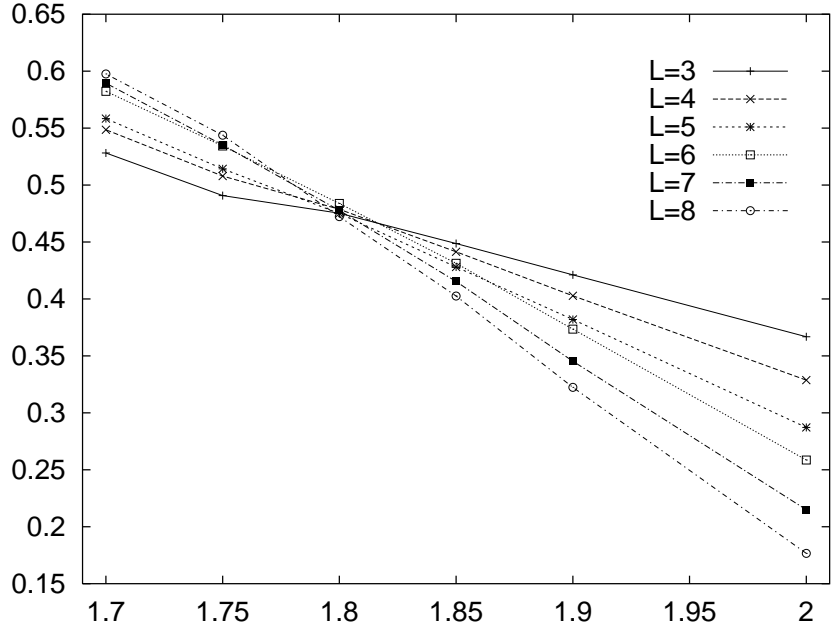,width=0.75\linewidth}
\end{center}
\protect\caption{$g(L,T)$ against $T$; the errors are of the
order of the symbol.  \protect\label{F_binder}}
\end{figure}

Defining the spin glass susceptibility as
\be
\chi_{SG}(L,T)=\frac1N\sum_{i,j}\overline{\lan\si\sj\ran^2}=Nq_L^{(2)}
\;\;,
\ee
where $N=L^4$, $\lan(\cdot\cdot)\ran$ is the thermodynamical average
and $\overline{(\cdot\cdot)}$ the mean over the disorder; and the
Binder parameter as
\be
g(L,T)=\frac12\left(3-\frac{q_L^{(4)}}{(q_L^{(2)})^2}\right) \;\; ,
\ee
the results of our simulations are plotted in fig.\ref{F_chi} for
the spin glass susceptibility and fig.\ref{F_binder} for the
Binder cumulant.
 
The errors on the plotted data are derived from a jackknife analysis,
which also confirms that the overlaps measured every $\tau$ sweeps are
decorrelated.
Using finite size scaling we see that $\chi_{SG}(L,T)$
and $ g(L,T)$ scale as (in the scaling region)
\be
\chi_{SG}(L,T)=L^{2-\eta}\tilde\chi_{SG}(L^{1/\nu}(T-\Tc)) \;\; ,
\ee
\be
g(L,T)=\tilde g(L^{1/\nu}(T-\Tc)) \;\; .
\ee
Note that at the critical temperature the Binder parameter does not
depend on the size of the system, so $\Tc$ is the temperature where
the curves of fig.\ref{F_binder} cross themselves.

In the neighborhood of \Tc\ we can approximate the function $\tilde g$
with a linear one and  obtain the following critical temperature
and $\nu$ exponent
\begin{eqnarray}
\Tc & = & 1.80 \pm 0.01 \;\; , \\
\nu & = & 0.9 \pm 0.1 \;\; .
\end{eqnarray}
The value of $\nu$ is also confirmed  by the results of the analysis
done, following \cite{THREE_REP}, on the quantity 
\be
\left. \frac{dg}{dT}\right|_{T_0:\; g(T_0)=g_0} = \alpha
L^{1/\nu} \;\; ,
\ee
obtaining
\be
\nu=1.06 \pm 0.06 \;\; .
\ee

The prediction about the infinite volume limit of the Binder cumulant
is different in the droplet theory
($g(L,T\!<\!\Tc)\!\stackrel{L\ra\infty}\lra\!1$) and in the SRSB
picture 
($g(L,T\!<\!\Tc)\!\stackrel{L\ra\infty}\lra \overline{g}(T)\!<\!1$).
Unfortunately with our data ($L=3$ to $L=8$) is impossible to
extrapolate the infinite value with a precision  that could
discriminate, without doubt, between the two predictions.

The estimation of the anomalous dimension $\eta$ can be performed by fitting
the $\chi_{SG}$ data, at $T=\Tc$, with a power law
\be
\chi_{SG}(L,T=\Tc) \propto L^{2-\eta} \;\; ,
\label{E_chi_scal}
\ee
obtaining $\eta = -0.35 \pm 0.05$ (the error is due mostly to the
indetermination on the critical temperature and to the rapid variation
in the region around \Tc\ of the exponent in eq.(\ref{E_chi_scal})).
These results are in agreement with those found by Bhatt and Young in
\cite{BHATT_YOUNG} using a maximum size of $6^4$ and 200-800 samples:
$\Tc = 1.75 \pm 0.05$, $\nu = 0.8 \pm 0.15$ and \mbox{$\eta = -0.3 \pm
0.15$}.

Using the scaling law $\gamma=\nu(2-\eta)$ and the exponents values
just calculated, we have $\gamma = 2.1 \pm 0.2$, which is in good
agreement with the value obtained by the high temperature expansions
$\gamma = 2.0 \pm 0.4$, \cite{HIGH_T_EXPAN}.

Another series of computer runs, performed using the annealing procedure
\cite{SIM_ANN}, let us measure the non-connected susceptibility for a
wide range of temperature in the spin glass phase ($T<\Tc$).
We clearly see that the data diverge with  increasing  system
sizes, even though, because of the small lattices, many different fits
are possible, e.g.
\be
\chi_{SG}(L,T) = A(T) L^4 \left[1+B(T) L^{-\Lambda(T)}\right] \;\;,
\ee
or 
\be
\chi_{SG}(L,T) \propto L^{2-\eta(T)} \;\; .
\ee
Further evidence for the value of \Tc\ can be obtained, as the highest
temperature where the power law fit is yet acceptable (by a $\chi^2$
test).

\section{\protect\label{S_dynamic}Off-equilibrium dynamics}

The second part of our study was devotes to the simulation of systems
of greater dimensions (ranging from $8^4$ to $32 \times 16^3$).
At the beginning of every simulation the system is frozen from an
infinite temperature to one in the critical region ($T<\Tc$), and
measures of the autocorrelation functions immediately start with the
system still out of equilibrium.
Due to the huges thermalization times of the simulated systems, the
off-equilibrium dynamics is the most realistic situation and also the
most interesting.
In fact, due to the enormous number of metastable states, the
dynamics is very slow and besides it is reminiscent of the time passed
in the spin glass phase, that we will call $t_w$.
These effects can be clearly seen by the study of the autocorrelation
functions defined as
\be
\label{E_aging}
C(t,t_w)=\frac1N\sum_{i=1}^N\overline{\lan\si(t_w)\si(t_w+t)\ran}\;\;,
\ee
where $\overline{(\cdot\cdot)}$ is the mean over the disorder and
$\lan(\cdot\cdot)\ran$ stands not for an average over the equilibrium
thermodynamic state, since we are not at the equilibrium, but for an
average over the thermal histories.
Nevertheless we found that, for the system sizes we considered,
disorder fluctuations are always stronger, so generally we omit the
angular brackets.

Our simulations cover the cold phase (from $T=\Tc=1.8$ down to
$T=0.2$) through the set of waiting times $t_w = 2^k$ with
$k=7,8,\ldots,21$ and averaging over 3072 disorder realizations
systems of volumes from $8^4$ to $12^4$.

In the four-dimensional Ising spin glass the presence of a critical
temperature and the subsequent spin glass phase has been widely
accepted, so the principal question that remain to answer is which
kind of phase space arises for $T<\Tc$.
In the literature there are principally two theories that try to
describe the spin glass systems in their low temperature phase: one is
based on a mean field like approximation which predicts a spontaneous
replica symmetry breaking (SRSB picture); the other one, starting
from a Migdal-Kadanoff renormalization group technique, concludes that
the system remains trivial, with only one pure state (droplet model).
The predictions of the two theories regarding the autocorrelation
function are different: the SRSB picture predicts that in the limit of
$t_w \ra \infty$ the autocorrelation must be a power law that
converge to the Edward-Anderson parameter ($q_{EA}$)
\be
C(t,t_w)=(q_{EA}+a t^{-x}) \frac{f(t/t_w)}{f(0)} \;\;,
\protect\label{E_corr_qea}
\ee
while in the droplet model the relaxation is slower
\be
C(t,t_w) = (\log t)^{-\theta/\psi}\: C'\left(
\frac{\log(t/\tau)}{\log(t_w/\tau)} \right) \;\;.
\protect\label{E_aging_droplet}
\ee
 
\begin{figure}
\begin{center}
\leavevmode
\centering\epsfig{file=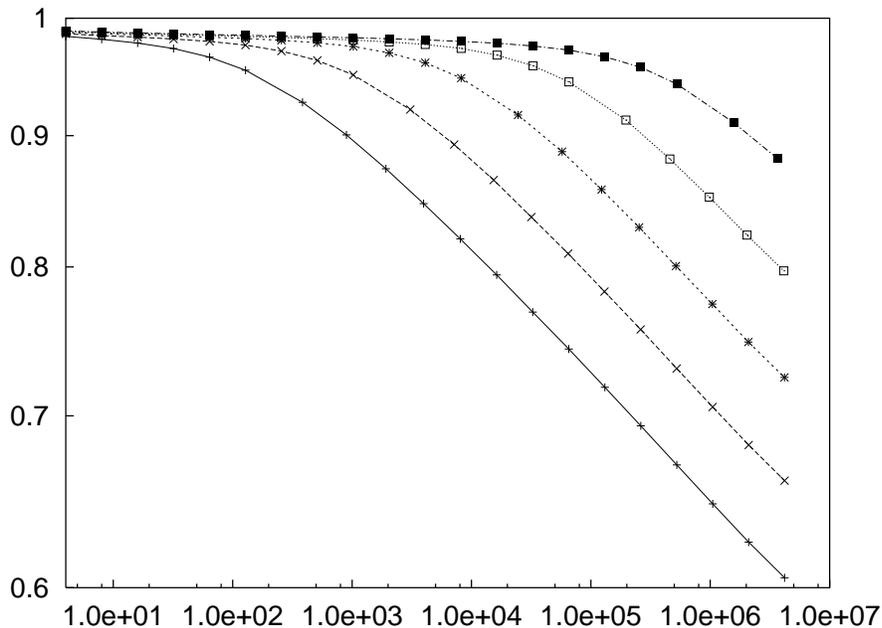,width=0.75\linewidth}
\end{center}
\protect\caption{$C(t,t_w)$ vs. $t$ at $T=0.2$ with $t_w = 2^7,
2^{10}, 2^{13}, 2^{16}, 2^{19}$ (bottom to top). 
\protect\label{F_aging_T020}} 
\end{figure}

\begin{figure}
\begin{center}
\leavevmode
\centering\epsfig{file=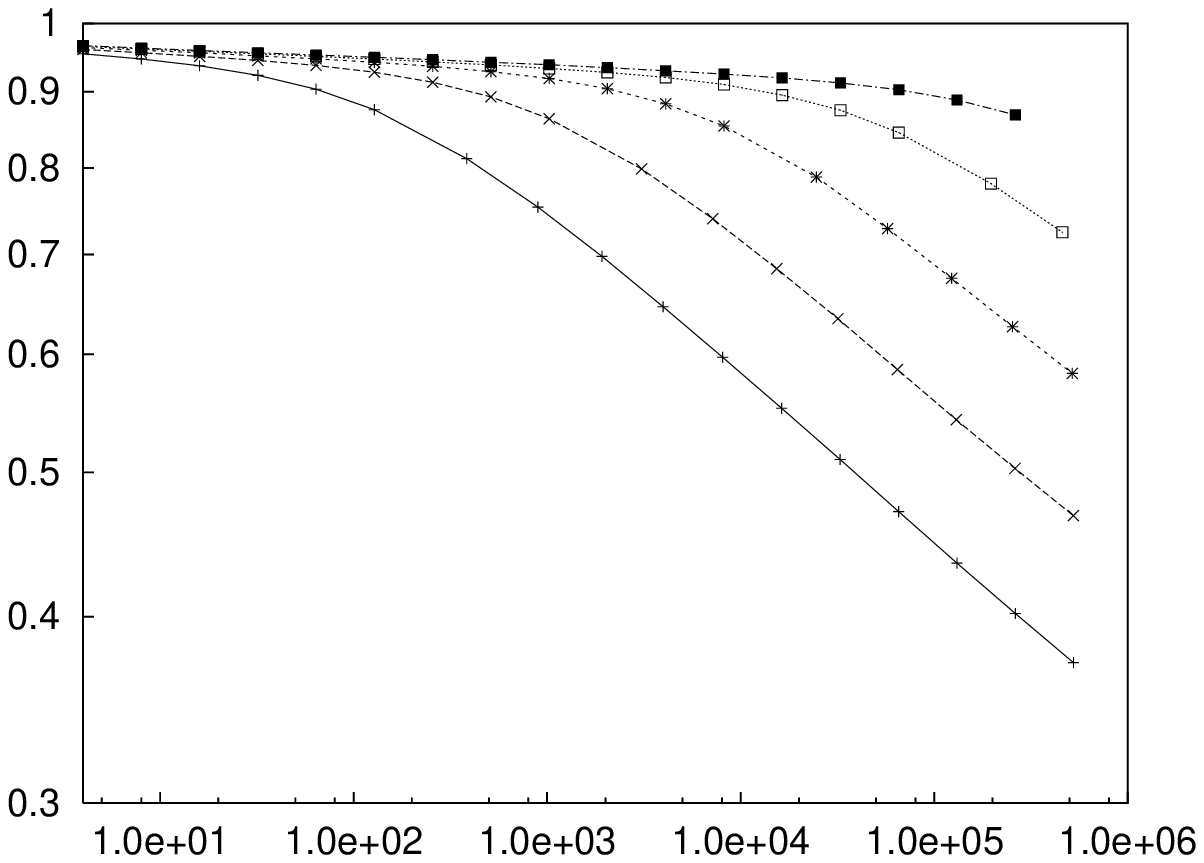,width=0.75\linewidth}
\end{center}
\protect\caption{$C(t,t_w)$ vs. $t$ at $T=0.45$ with $t_w = 2^7,
2^{10}, 2^{13}, 2^{16}, 2^{19}$ (bottom to top). 
\protect\label{F_aging_T045}}
\end{figure}

\begin{figure}
\begin{center}
\leavevmode
\centering\epsfig{file=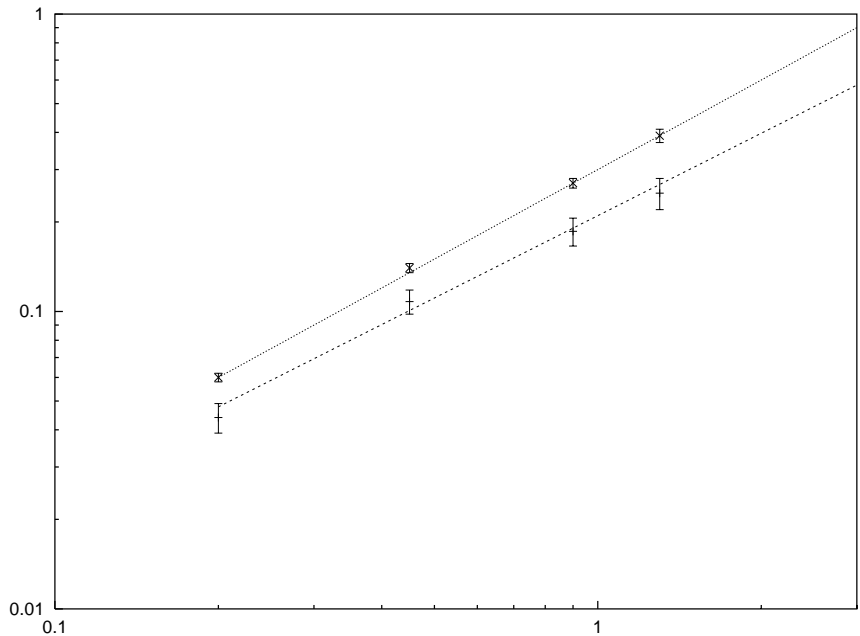,width=0.75\linewidth}
\end{center}
\protect\caption{$\lambda_{t_w=0}(T)$ (top) and $\lambda_{t_w\ne
0}(T)$ (bottom) against $T$; the lower line represents the power fit
$\lambda(T)=0.21(1)\,T^{0.92(7)}$; the upper line both the linear and
the power fit: respectively $\lambda(T)=0.000(3)+0.30(1)\,T$ and
$\lambda(T)=0.303(8)\,T^{1.00(3)}$. \protect\label{F_lambda_T}}
\end{figure}

\begin{figure}
\begin{center}
\leavevmode
\centering\epsfig{file=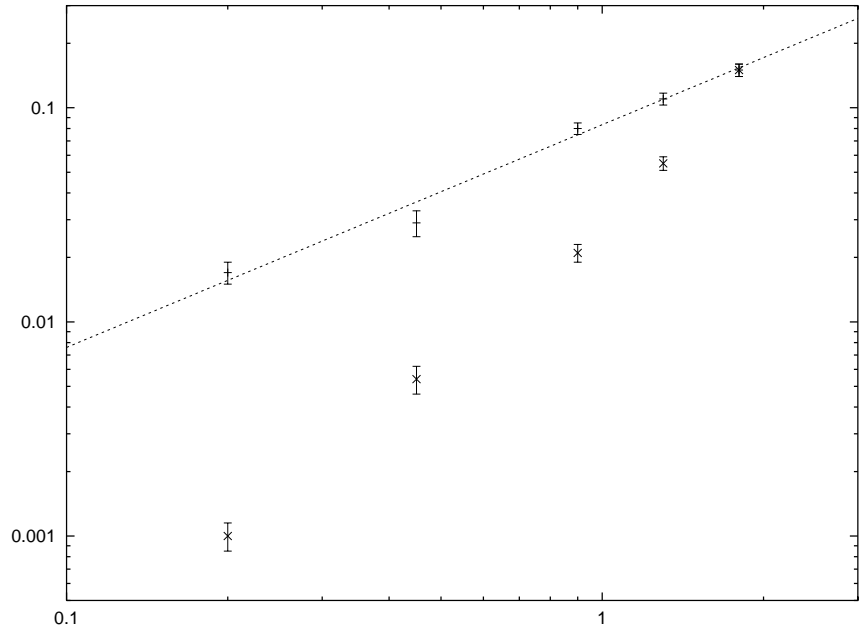,width=0.75\linewidth}
\end{center}
\protect\caption{$x(T)$ (top) and $x'(T)$ (bottom) against $T$ (the
value at the greater temperature is equal: $x(\Tc)=x'(\Tc)$); the line
is best power fit $x(T)=0.083(3)\,T^{1.04(7)}$. \protect\label{F_x_T}}
\end{figure}

The data we collected (see fig.\ref{F_aging_T020} and
fig.\ref{F_aging_T045}) seem to agree with the scaling law used in
\cite{RIEGER_93} and in \cite{CU_KU_RI} 
\be
C(t,t_w)=t^{-x'(T)}\tilde C(t/t_w) \;\;,
\protect\label{E_aging_srsb}
\ee
with a scaling function
\be
\tilde C(z)=\left\{
\begin{array}{ll}
{\rm constant} & {\rm for} \;\; z \rightarrow 0 \\
z^{x'(T)-\lambda(T)} & {\rm for} \;\; z \rightarrow \infty
\end{array}
\right.
\ee
The values of the exponent $x'(T)$  is plotted together with the
values of $x(T)$ in fig.\ref{F_x_T}, while $\lambda(T)$ is plotted in
fig.\ref{F_lambda_T} .

\begin{figure}
\begin{center}
\leavevmode
\centering\epsfig{file=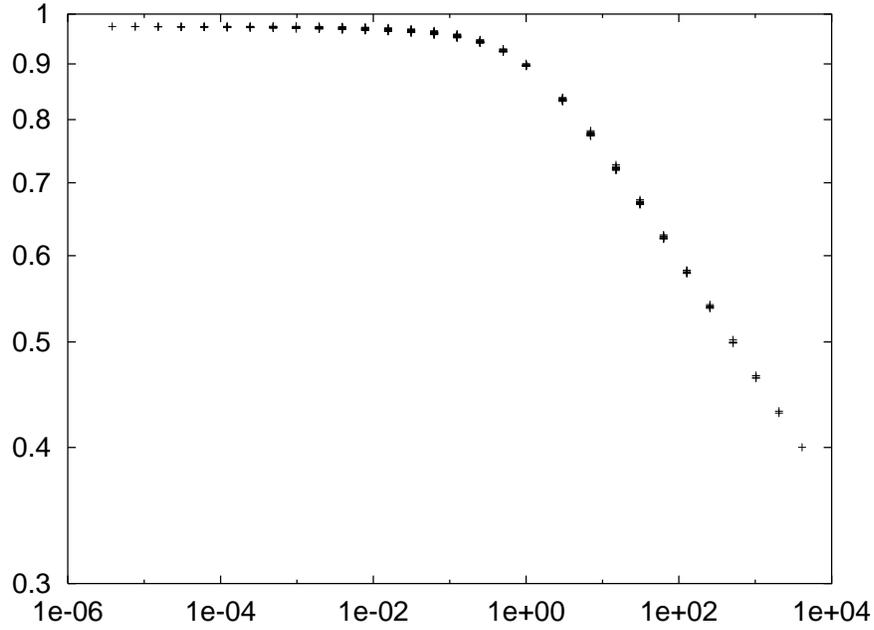,width=0.75\linewidth}
\end{center}
\protect\caption{$T=0.45$ aging autocorrelation function rescaled
following eq.(\protect\ref{E_aging_srsb}), with \mbox{$x=0.0054$.} We
plot $t^x C(t,t_w)$ against $t/t_w$. \protect\label{F_scaled_srsb}} 
\end{figure}

\begin{figure}
\begin{center}
\leavevmode
\epsfig{file=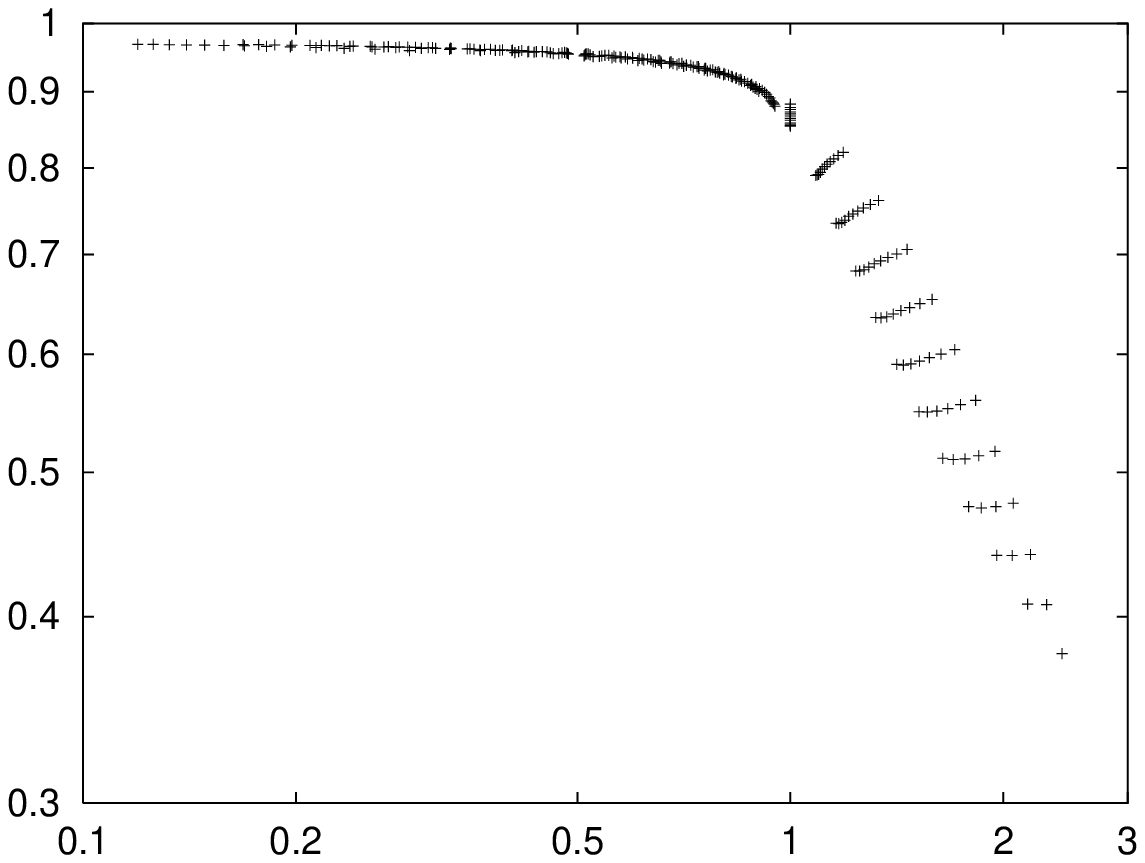,width=0.45\linewidth}
\epsfig{file=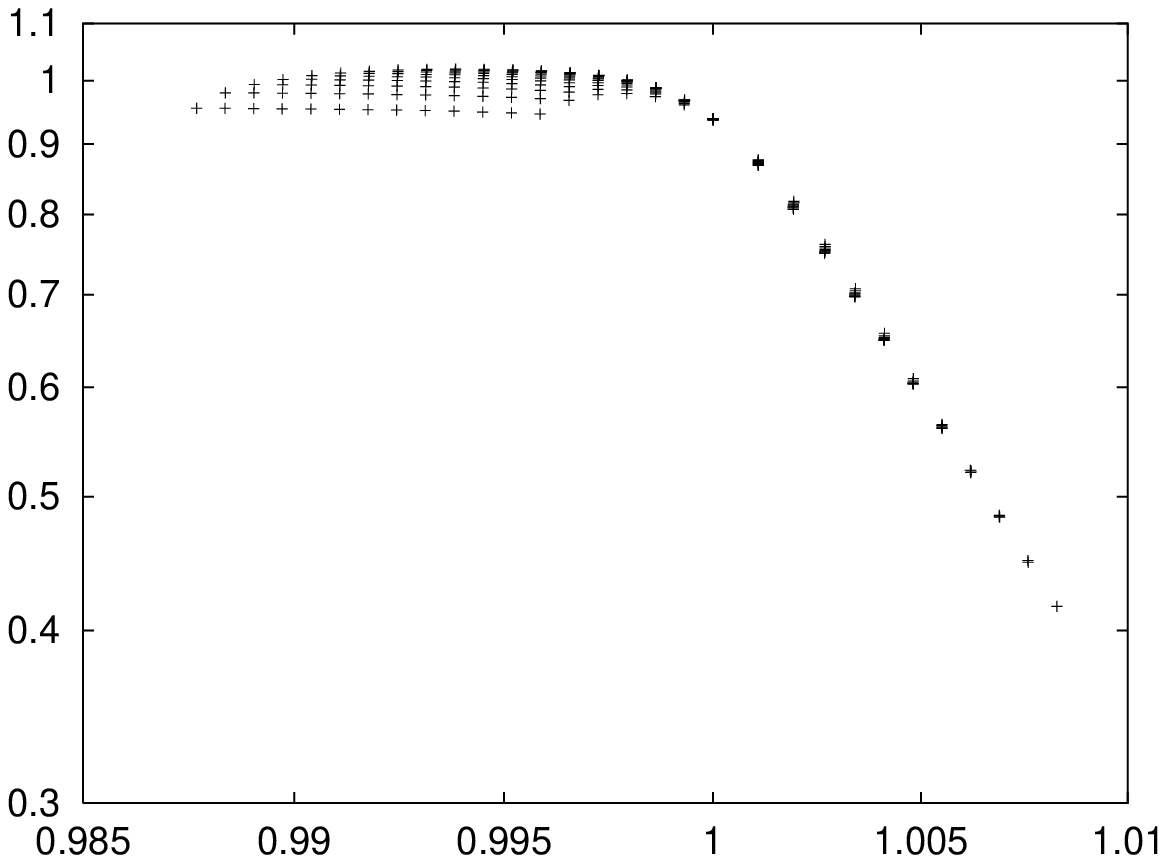,width=0.45\linewidth}
\end{center}
\protect\caption{Tentatives of rescaling following the droplet model
law eq.(\protect\ref{E_aging_droplet}): in the left plot
$\theta/\psi=0.0054$ and $\log(\tau)=-1$; in the right plot
$\theta/\psi=0.043$ and $\log(\tau)=-1000$.
 \protect\label{F_scaled_droplet}} 
\end{figure}

To evaluate the goodness of the two proposed scaling formulas 
eq.(\ref{E_aging_srsb}) and eq.(\ref{E_aging_droplet}) we plotted in
fig.\ref{F_scaled_srsb} the $T=0.45$ data rescaled with the former
law, noting that they collapse very well on a single curve.
On the contrary, using the droplet model scaling law, it was
impossible for us to find a value for the parameters $\theta/\psi$ and
$\tau$ such to force the data over a single curve; in
fig.\ref{F_scaled_droplet} we show the best scalings we could
obtain in order to make the data collapse in the $t<t_w$ or in the
$t>t_w$ region.

Nevertheless the very good rescaling of the data in
fig.\ref{F_scaled_srsb} we also performed a deeper analysis in order
to find the value of the Edward-Anderson parameter, $q_{EA}$, which is
assumed to be zero in eq.(\ref{E_aging_srsb}).
The value of $q_{EA}$ can be found performing the $t \ra \infty$ limit
{\em after} the $t_w \ra \infty$ limit; for this purpose we have done
very long simulations (more than 4 millions Monte Carlo sweeps).
We note that the scaling laws obeyed by the data in the two regions
$t \gg t_w$ and $t \ll t_w$ are essentially different.
In the former the data can be fitted by a power law of the ratio 
$t/t_w$, while in the latter we obtain a law equal to that of
eq.(\ref{E_corr_qea}) times a function of $t/t_w$ which is almost a
constant.

Such a behaviour for $C(t,t_w)$ can be justified assuming that the
system evolves as long as $t\ll t_w$ with a quasi equilibrium dynamics 
which converges to $q_{EA}$ while for $t \gg t_w$ it decorrelates
faster and toward zero ($C\sim t^{-\lambda}$ with $\lambda(T) \gg x(T)
\;\; \forall T$), but always with a critical slowing-down.

The values for $\lambda(T)$ have been obtained fitting the
$C(t,t_w=0)$ data with a power law and for $t_w \neq 0$ with 
\be
C(t,t_w) \propto \left(\frac{t}{t_w}\right)^{-\lambda(T)} \;\;,
\ee
in the range $t/t_w \ge 15$. In fig.\ref{F_lambda_T} we plot the
results either for $t_w=0$ and for $t_w \ne 0$.
We note that both  fits are compatible with the linear dependence
in the temperature predicted from the experimental measurements
\cite{EXPER}.

\begin{figure}
\begin{center}
\leavevmode
\centering\epsfig{file=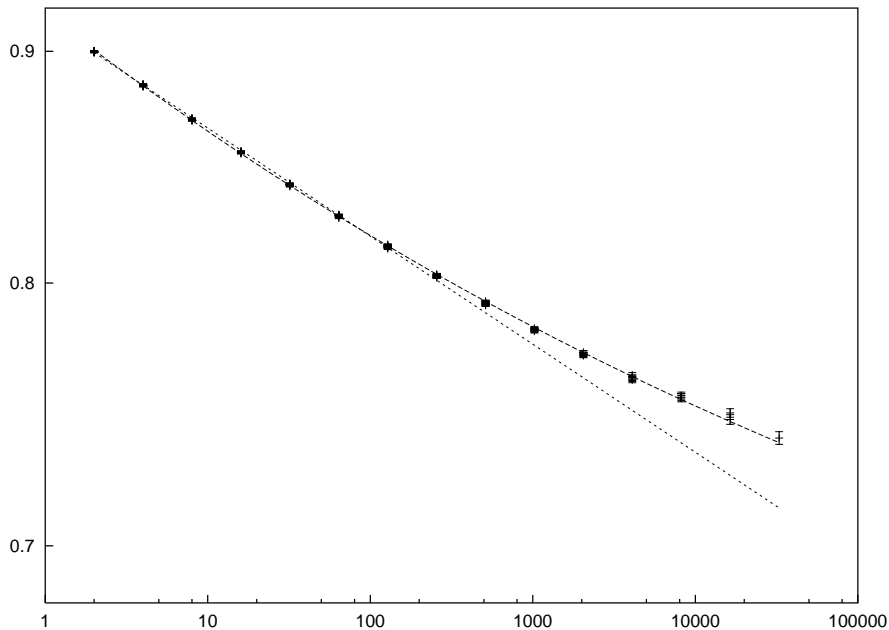,width=0.75\linewidth}
\end{center}
\protect\caption{${\displaystyle\frac{C(t,t_w)}{\overline{C}(t/t_w)}}$
data at $T=0.9$ versus $t$; the upper line is the best power plus
constant fit: $0.60(4)+0.32(4)\,t^{-0.08(1)}$, while the lower line is
the best power fit. \protect\label{F_rescaled_090}}
\end{figure}

In the region $t_w/t \ge 32$ we have performed the analysis assuming
that the correlation function could be factorized as
\be
C(t,t_w) = (q_{EA}+a t^{-x}) \overline{C}(t/t_w) \;\;,
\ee
where we have approximated $\overline{C}(z)=1-c_1 z^{c_2}$ for $z \ra
0$.
First of all, the rescaling function $\overline{C}(t/t_w)$ has been
calculated fitting the correlation function at a fixed value of $t$.
Later, once divided the data by this function, we verified that the
curves for differents ratios $t/t_w$ collapse over a single curve and
we interpolated the data via a power law plus constant, following
eq.(\ref{E_corr_qea}).
In fig.\ref{F_rescaled_090} we plot in a log-log scale typical
$C(t,t_w)/\overline{C}(t/t_w)$ data with the best fit; we note that up
to now in the literature this data have been fitted via a simple power
law, while it is evident that the points in fig.\ref{F_rescaled_090}
are not on a straight line.

\begin{figure}
\begin{center}
\leavevmode
\centering\epsfig{file=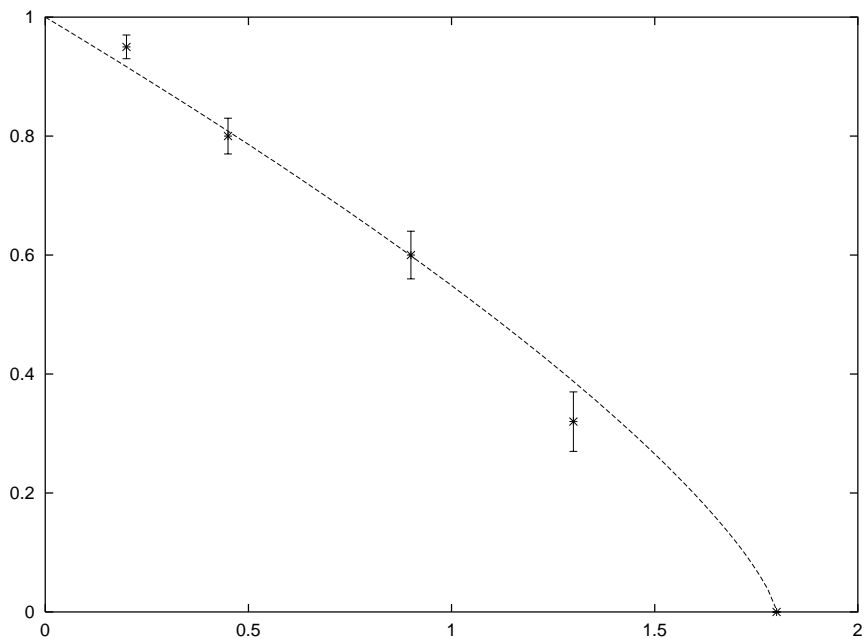,width=0.75\linewidth}
\end{center}
\protect\caption{Edward-Anderson order parameter against $T$; the line
is only a guide to the eye as explained in the text.
 \protect\label{F_qEA_T}}
\end{figure}

From these fits the values of $q_{EA}$ and $x$ as a function of the
temperature (see fig.\ref{F_x_T} and fig.\ref{F_qEA_T}) can be
obtained. As a guide to the eye, we plot in fig.\ref{F_qEA_T} the
simpler function that behaves like $|T-\Tc|^\beta$ near the critical 
temperature and tends to 1 for $T=0$
\be
q_{EA}(T)=\left(\frac{\Tc - T}{\Tc}\right)^\beta \;\;,
\ee
where $\Tc=1.8$ and $\beta=\frac{\nu}{2}(d-2+\eta)=0.74$ (using the
values found in the previous section).
From fig.\ref{F_x_T} we note that only the quantity $x(T)$, and  not
$x'(T)$, is such that $x(T)/T$ is roughly independent from the
temperature, so that only in this parametrization the $t_w=\infty$
autocorrelation function ($R(t;T)=C(t,t_w=\infty)$ at temperature $T$)
can be written as
\be
R(t;T) - R(\infty;T) = b(T) \exp(-B\,T \log(t)) \;\;.
\ee
The relevance of the variable $T\log(t)$ has been observed in
experiments on magnetic remanence in a wide region \cite{EXPER}.

We call off-equilibrium correlation length, $\xi(t)$, the typical
distance over which the system is thermalized after a time $t$.
For this domain growth the SRSB picture predicts a power law 
\cite{MA_PA_RI_RU_95}
\be
\xi(t) \propto t^{1/z(T)} \;\;,
\ee
while in the droplet model, where the energy barriers scale
proportionally to $L^\psi$, the law is
\be
\xi(t) \propto (T \log t)^{1/\psi} \;\;.
\ee

At least at the critical temperature there is a scaling relationship
between the dynamical exponent $z(\Tc)$ and the one which describes the
dynamics in the quasi equilibrium regime, $x(\Tc)=x'(\Tc)$ (because
$q_{EA}(T=\Tc)=0$)
\be
x = \frac{d-2+\eta}{2z} \;\;.
\ee
This equation is satisfied by all the exponents we have estimated at
the critical temperature: $x=0.15$, $\eta=-0.35$ and $z=5.3$.

To find the behaviour of the off-equilibrium correlation length we
have measured, like in \cite{MA_PA_RI_RU_95}, the equal time spatial
correlation functions
\be
G(r,t) = \frac1N \sum_{i=1}^N \overline{\lan \si(t) \sigma_{i+r}(t)
\ran^2} \;\;,
\ee
where the averages are the same as in eq.(\ref{E_aging}) and $t$ is
the time since the cooling.
This study has been performed on systems of volume $32 \times 16^3$.

\begin{figure}
\begin{center}
\leavevmode
\centering\epsfig{file=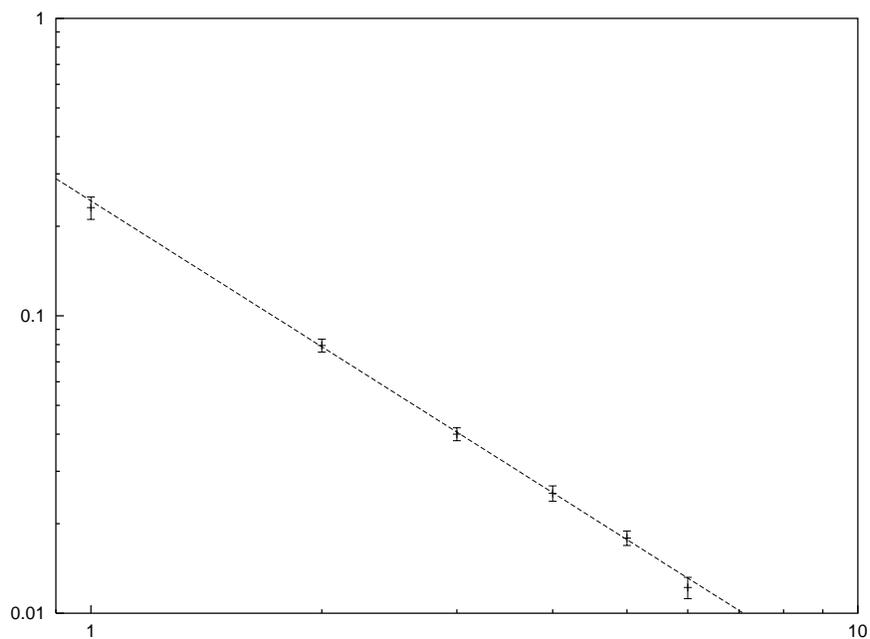,width=0.75\linewidth}
\end{center}
\protect\caption{$G(r,t=\infty)$ vs. $r$ at $T=\Tc=1.8$; the line is
the power fit $0.24(1)\: r^{-1.63(5)}$. \protect\label{F_space_corr}}
\end{figure}

From scaling concepts we know that, at large values of $r$, $G(r,t)$
must behave like
\be
G(r,t) \propto r^{-(d-2+\eta)} f\left( \frac{r}{\xi(t)} \right) \;\;,
\ee
and, supposing $f(y)=A \exp(-B y^D)$, we have fitted our data with the
function
\be
G(r,t)=Ar^{-(2+\eta)}\exp\left[-B\left(\frac{r}{t^{1/z}}\right)^D\right]
\;\;.
\ee
In the $t\ra\infty$ limit the exponential term tends to 1 and we
obtain a spatial correlation function that decrease with a power law:
in fig.\ref{F_space_corr} we plot such function at the critical
temperature ($\Tc=1.8$).
Note that from the slope of the curve we obtain an estimation of the
$\eta$ exponent compatible with that of section \ref{S_static}.
 
\begin{figure}
\begin{center}
\leavevmode
\centering\epsfig{file=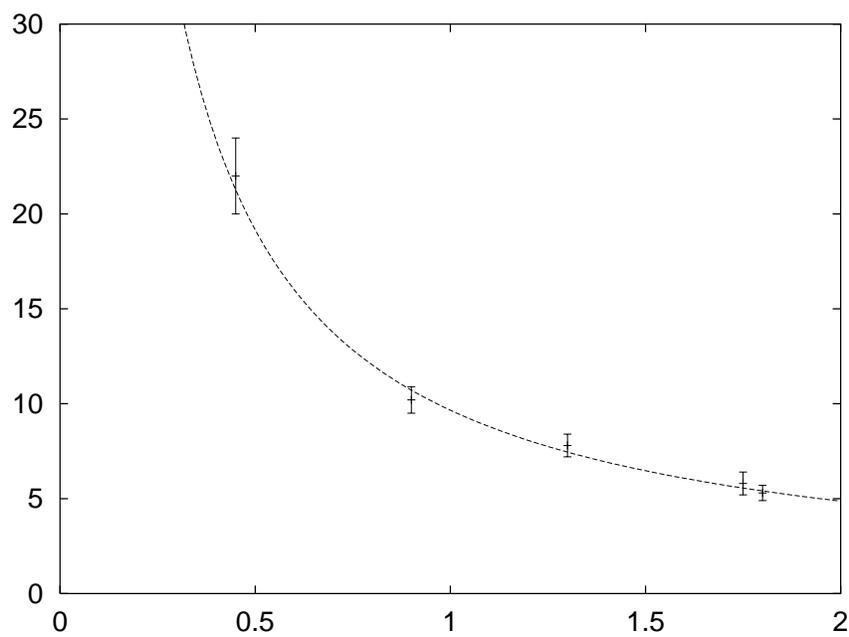,width=0.75\linewidth}
\end{center}
\protect\caption{$z(T)$ vs. $T$; the line is the power
fit, eq.(\protect\ref{E_z_T}). \protect\label{F_z_T}}
\end{figure}

At the lower temperatures the value of $\eta$ strongly  depends on the
$r$ range of interpolation, because the fitting function diverges at
$r=0$.
On the contrary, trying to fit the data in different ranges of $r$, we
find that the dynamical exponent $z(T)$ is a robust parameter which
remains unchanged for every $r$ range (plotted in fig.\ref{F_z_T})
 
Fitting the plotted data with a power law we obtain, up to the
critical temperature
\be
\label{E_z_T}
z(T) = A\ T^{-\alpha} \;\;,
\ee
with $A = 9.7 \pm 0.5$ and $\alpha = 1.0 \pm 0.1$.

A preliminary analysis of a new set of data at $T=0.9=\Tc/2$ suggests
a value of $\eta \simeq -1$.
The fact that the value of $z$ at this temperature is higher than the
corresponding value at \Tc\ makes the evaluation of the exponent $\eta$
prone to systematic error.
Nevertheless this rough estimate of $\eta$ is compatible with the
prediction of the reference \cite{KONDOR}.

\section{\protect\label{S_CONCLU}Conclusions}

In this paper we have studied the on and off-equilibrium properties of
the four dimensional Gaussian spin glass. 

In the static case the hypothesis of a transition {\it \`a la}
Kosterlitz-Thouless have been rejected owing to the study of the
on-equilibrium properties of the model: the existence of a second
order phase transition is well testified by the clean cut of the
Binder cumulant curves.
We have determined with more precision that in previous simulations
both the critical temperature as well as the critical exponents.

In the off-equilibrium case we have settled, for the first time, a
form of the autocorrelation function compatible, in the large times
limit ({\it i.e.} on equilibrium), with the existence of an order
parameter different from zero.
We have been able to determine, in a dynamical way, the value of
$q_{EA}$ as a function of the temperature (this again confirm the
absence of a Kosterlitz-Thouless transition).
Also we have established the temperature dependence of the exponents
that appear in these off-equilibrium dynamics, linear in all the
cases.

The dynamics of the model seem to be described better by the SRSB
theory than by the droplet theory: in effect the autocorrelation
functions, properly rescaled, follow very well the power laws
predicted by the former, while are quite different from the
logarithmic laws predicted by the latter.

Our conclusion is that the SRSB theory seems to be, at the moment, the
best picture to describe the EA model in finite dimensions greater
than the lower critical dimension.

A still open problem we are planning to study in the future regards
the estimate, simulating much larger lattices, of the order parameter
using the static spin glass susceptibility.

\section{\protect\label{S_ACKNOWLEDGES}Acknowledgments}

We acknowledge useful discussions with E. Marinari.
We are very grateful to the APE group for his continuous support and
valid assistance.
J. J. Ruiz-Lorenzo is supported by an EC HMC(ERBFMBICT950429) grant.

\end{document}